\documentclass[preprint]{elsarticle}
\usepackage{adjustbox}
\usepackage{subcaption}
\usepackage{xcolor}
\usepackage{lineno,hyperref}
\usepackage[ruled,vlined]{algorithm2e}
\usepackage{dingbat}
\usepackage{amsmath,amssymb}
\usepackage{multirow}
\DeclareMathOperator{\E}{\mathbb{E}}
\usepackage[ruled,vlined]{algorithm2e}
\modulolinenumbers[5]

\journal{Computers in Biology and Medicine}
\usepackage{etoolbox}
\makeatletter
\patchcmd{\ps@pprintTitle}
  {Preprint submitted to Computers in Biology and Medicine}
\makeatother

\begin{document}

\begin{frontmatter}

\title{MITS-GAN: Safeguarding Medical Imaging from Tampering with Generative Adversarial Networks}

\author[1]{Giovanni Pasqualino} 
\author[1]{Luca Guarnera} 
\author[1]{Alessandro Ortis\corref{cor1}}
\cortext[cor1]{Corresponding author:}
\ead{alessandro.ortis@unict.it}
\author[1]{Sebastiano Battiato}

\address[1]{Department of Mathematics and Computer Science, University of Catania, Viale Andrea Doria 6, Catania, 95125, Italy}

\begin{abstract}
The progress in generative models, particularly Generative Adversarial Networks (GANs), opened new possibilities for image generation but raised concerns about potential malicious uses, especially in sensitive areas like medical imaging. This study introduces MITS-GAN, a novel approach to prevent tampering in medical images, with a specific focus on CT scans. The approach disrupts the output of the attacker's CT-GAN architecture by introducing finely tuned perturbations that are imperceptible to the human eye.
Specifically, the proposed approach involves the introduction of appropriate Gaussian noise to the input as a protective measure against various attacks. Our method aims to enhance tamper resistance, comparing favorably to existing techniques.
Experimental results on a CT scan demonstrate MITS-GAN's superior performance, emphasizing its ability to generate tamper-resistant images with negligible artifacts. As image tampering in medical domains poses life-threatening risks, our proactive approach contributes to the responsible and ethical use of generative models. This work provides a foundation for future research in countering cyber threats in medical imaging.
Models and codes are publicly available~\footnote{https://iplab.dmi.unict.it/MITS-GAN-2024/}.
\end{abstract}


\end{frontmatter}

\linenumbers
\nolinenumbers
\section{Introduction}
\begin{figure}[t!]
\centering
\includegraphics[width=0.7\textwidth]{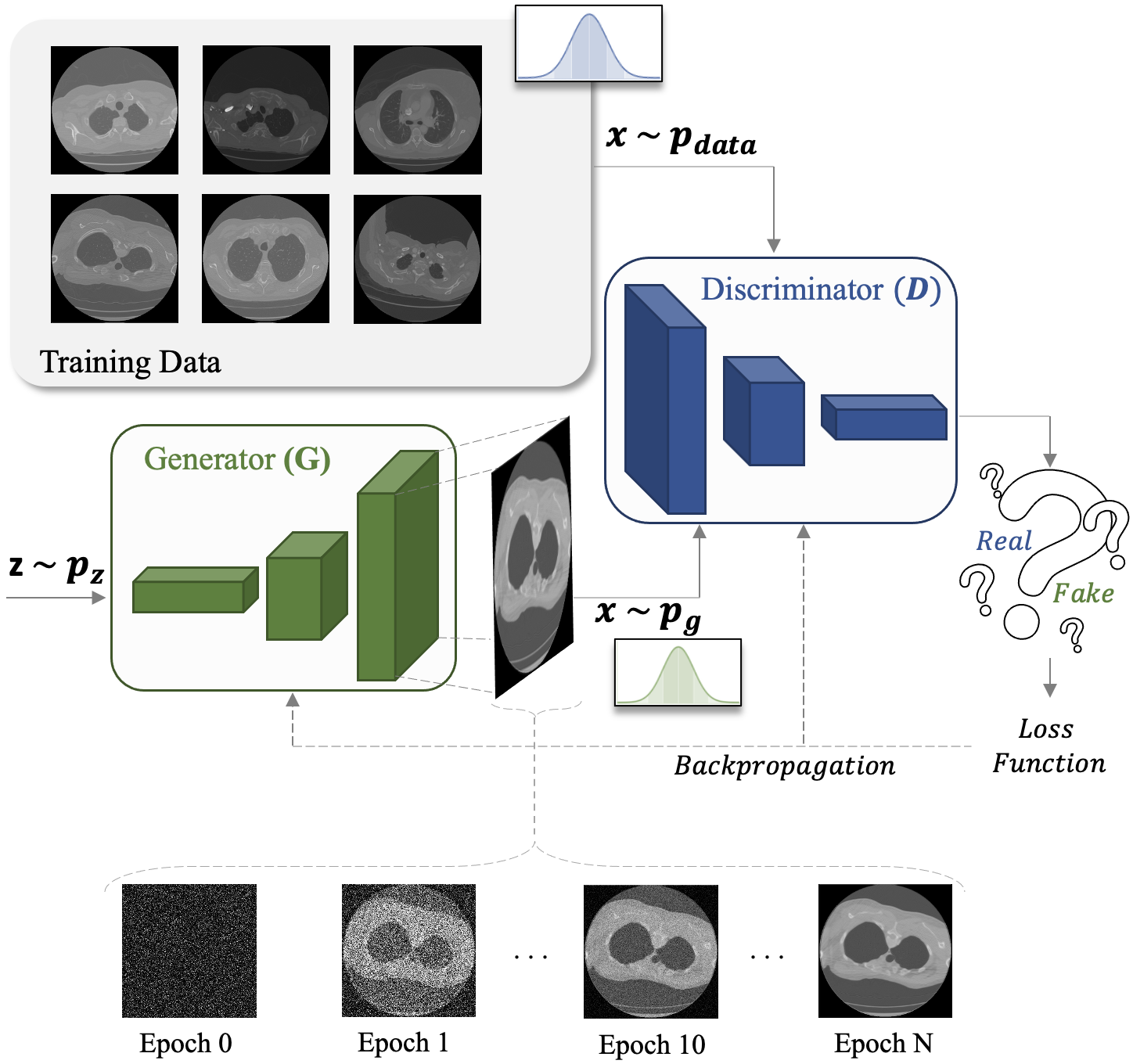}
\caption{Overview of the GAN architecture and training process.}
\label{fig:gan}
\end{figure}
In recent years, advancements in generative models have ushered in a new era of image generation and manipulation, showcasing remarkable capabilities in rendering images increasingly indistinguishable from their original counterparts~\cite{CycleGAN2017, brock2018large, choi2020stargan}. This progress, driven by deep learning techniques, has found applications in various domains, from creative artistry~\cite{shahriar2022gan} to medical imaging~\cite{madani2018chest}, among others.

In medical imaging, GANs have been instrumental in addressing the challenge of data scarcity. They are used to augment datasets by generating synthetic images or translating images between different modalities. For instance, GANs have been employed to convert MRI images into CT images~\cite{jin2019deep}, generate realistic 2D brain MRI images~\cite{bermudez2018learning}, and even enhance image resolution~\cite{gupta2020super}. These applications not only improve the quality and availability of medical images but also support advancements in diagnostic processes.
Islam et al.~\cite{islam2020gan} proposed a GAN-based method to generate PET images of the brain. This new dataset could be used to create new artificial intelligence methods to help doctors make an early diagnosis of Alzheimer's disease.
Due to the absence of Arterial Spin Labelling (ASL) data, Li et al.~\cite{li2021new} proposed a GAN architecture in order to synthesise such images. ASL measures cerebral blood flow, which is useful for making diagnoses for dementia diseases.
Pang et al.~\cite{pang2021semi} proposed a semi-supervised GAN architecture to perform data augmentation on ‘breast ultrasound mass’ images, in order to significantly improve the performance of the TCGAN classifier, created to discriminate the presence or absence of breast cancer.
Liu et al.~\cite{liu2021ct} proposed a multi-cycle GAN to generate CT images from MRI images, overcoming the limitations of MRI in that no information about the patient's bones is obtained. The technique reduces patients' exposure to radiation, improving the safety of radiotherapy. In general, MRI images contain noise that can be removed with the conditional GAN proposed by Tian et al.~\cite{tian2021boosting}. This work exceeds state-of-the-art methods in both noise reduction and the preservation of robust anatomical structures and defined contrast.
A very interesting approach was proposed by Dong et al. \cite{dong2019automatic}, in which a GAN architecture was used to automatically segment CT images of the thorax, using a U-Net architecture as generator and FCN as discriminator, in order to improve radiotherapy treatment planning. The proposed architecture achieved better segmentation results than state-of-the-art approaches.
However, alongside positive applications, researchers have demonstrated the malicious use of GANs~\cite{goodfellow2014generative} for tasks such as malware obfuscation~\cite{hu2022generating} and the creation of deepfakes~\cite{chesney2019deep}.
The key idea behind GANs involves training two neural networks, a generator, and a discriminator, in an adversarial setting. The generator aims to produce synthetic data, such as images, that is indistinguishable from real data, while the discriminator's task is to differentiate between real and generated data. This adversarial training process results in the generator continually improving its ability to create realistic data, making GANs highly effective in image generation tasks (Figure~\ref{fig:gan}).
\begin{figure}[t!]
\centering
\includegraphics[width=0.7\textwidth]{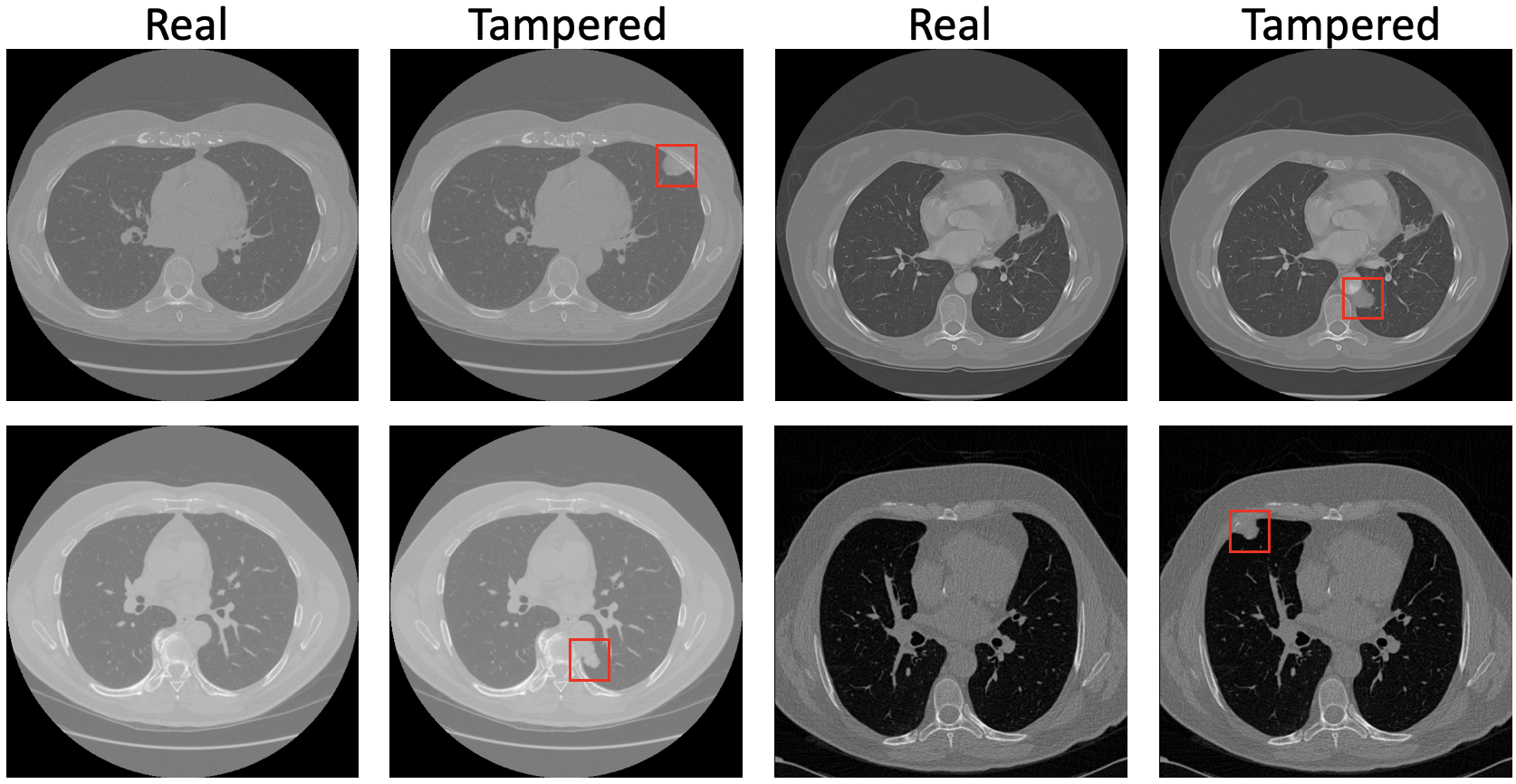}
\caption{Qualitative results comparison between real and tampered CT scans. Columns 1 and 3 show the original images, whereas Columns 2 and 4 depict the manipulated images. The red bounding boxes highlight the manipulations introduced by CT-GAN, wherein tumors have been added to the scans. This visual representation underscores the impact and detectability of manipulations within the medical imaging context.}
\label{fig:medical_img}
\end{figure}
Within the medical domain, the potential consequences of malicious tampering are critical, as the integrity and authenticity of images can have life-or-death implications as shown in Figure~\ref{fig:medical_img} manipulating the images provided by the authors of~\cite{chesney2019deep}. Image tampering techniques~\cite{mirsky2019ct} have raised concerns by highlighting the potential for malicious manipulation of medical images, such as computed tomography (CT) scans and radiographs. This introduces a new dimension of cyber attacks, with image manipulation being employed to deceive medical professionals and compromise patient care, potentially leading to misdiagnoses. To address this challenge, the research community has focused on developing automated detection systems for image manipulation, treating it as a classification task. Various learning-based approaches have shown promise, achieving excellent classification accuracy~\cite{cozzolino2021id, li2020face, nguyen2019capsule, marra2018detection}. Alternatively, another strategy is to prevent manipulations at the source by disrupting manipulation methods' output~\cite{yeh2020disrupting, ruiz2020disrupting, aneja2022tafim}. The key idea is to disrupt generative neural network models by introducing noise patterns at a low level, making it more challenging for malicious actors to create convincing forgeries.
\begin{figure}[t!]
\centering
\includegraphics[width=0.7\textwidth]{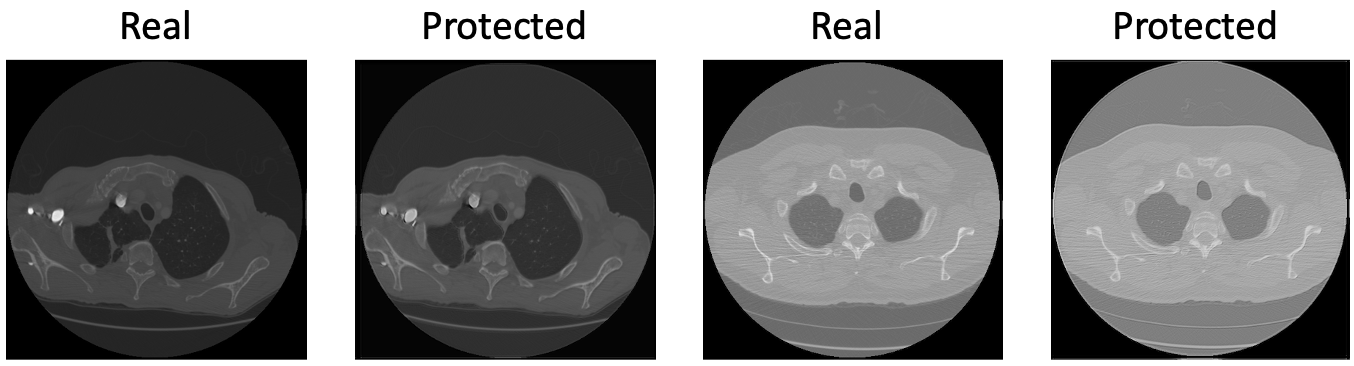}
\caption{Comparison between Real unprotected CT scans and protected CT scans generated by the proposed model MITS-GAN. As can be noted, the protected images, which embed the protection noise pattern, are similar to the original one.}
\label{fig:real_protected_comparison}
\end{figure}
In this study, we investigate the problem of image tampering in the medical domain, focusing on the manipulation of CT scans. 
To address this problem, we propose \textit{Medical Imaging Tamper Safe-GAN} (MITS-GAN) method. In particular, we introduce a framework based on 
Generative Adversarial Networks with the aim to generate protected images against image manipulation model~\cite{mirsky2019ct}. Our model generates the protected scans introducing an imperceptible noise with the aim to disrupt the output when the manipulation is performed and minimizing potential artifact that could pose challenges during the review process by medical experts (Figure~\ref{fig:real_protected_comparison}). MITS-GAN is designed to protect medical images from tampering, addressing risks such as misdiagnosis and medical fraud. Real-world concerns include manipulating CT scans to deceive doctors or commit insurance fraud, as well as using deepfake technology to fabricate medical images.
This research is significant for its potential to enhance diagnostic accuracy and bolster healthcare cybersecurity. By ensuring the authenticity of medical images, MITS-GAN supports reliable diagnoses, safeguards patient data, and prevents the misuse of AI technologies in healthcare.

The main contributions of the proposed work are:
\begin{itemize}
\item We address the critical issue of medical image tampering by proposing a robust methodology that ensures the integrity and reliability of diagnostic images. This approach is motivated by the urgent need to protect medical imaging from manipulation, which could otherwise compromise diagnostic accuracy and the reliability of Machine Learning methods and other systems based on such datasets;
\item We introduce a novel framework called MITS-GAN (Medical Imaging Tamper Safe-GAN) and compare its performance with state-of-the-art methods. MITS-GAN leverages Generative Adversarial Networks (GANs) to safeguard medical images from tampering. Our results demonstrate the superior effectiveness of MITS-GAN in preserving the authenticity and reliability of medical images;
\item Our work lays the groundwork for future research aimed at mitigating cyber threats in the field of medical imaging. We emphasize the importance of proactive measures to protect and maintain the integrity of medical scans, highlighting the long-term implications of our approach for the security of medical data.
\end{itemize}

The document is organized as follows. Section \ref{sec:sota} reports the main works in the literature. 
The proposed approach is described in Section \ref{sec:proposed}. The  used for the experiments, the metrics to evaluate the performances, the experimental results and comparison are reported in Section~\ref{sec:dataset_metrics}. Finally, Sections~\ref{sec:discussion} and \ref{sec:conclusion} conclude the paper with some hints for future works.

\section{Related Work}
\label{sec:sota}
\subsection{GAN Applications in Medical Imaging}
GANs have made significant contributions to the field of medical imaging, addressing various challenges and enhancing the quality and accessibility of medical imagery. GANs' ability to generate realistic images has been leveraged to alleviate the common issue of data scarcity in medical image analysis by augmenting s through the generation of new images or style translation. For instance, the authors of~\cite{bi2017synthesis} utilized a conditional GAN (cGAN) to transform 2D slices of CT images into PET images. The authors of~\cite{ben2017virtual} and~\cite{ben2019cross} demonstrated a similar approach employing a fully convolutional network with a cGAN architecture. In~\cite{dou2018unsupervised}, domain adaptation was employed to convert MRI images into CT images, while the authors of~\cite{jin2019deep} used CycleGAN to convert MRI images into CT images and vice versa. The authors of~\cite{bermudez2018learning}  use a deep convolutional GAN (DCGAN) to generate 2D brain MRI images. In~\cite{frid2018gan}, the authors used a DCGAN to generate 2D liver lesions. In~\cite{wolterink2018blood}, the authors generated 3D blood vessels using a Wasserstien (WGAN). In~\cite{madani2018chest}, the authors train two DCGANs for generating 2D chest X-rays (one for malign and the other for benign). 
Within the medical imaging domain, GANs have also found other interesting applications in segmentation~\cite{han2018spine}, super-resolution~\cite{gupta2020super} and anomaly detection~\cite{schlegl2017unsupervised}.

\subsection{Deepfake Detection Methods}
The ability to understand if an image is generated by a generative Neural Network is in some case challenging also for the human eyes representing a complicated problem.  To address this problem, numerous methods have been developed over the years to determine the authenticity of an image~\cite{masood2023deepfakes}. 

Researchers have demonstrated that generative engines leave traces on synthetic content that can be detected in the frequency domain~\cite{guarnera2020preliminary,Durall_CVPR_2020}. 
Giudice et al.~\cite{Giudice_JI_2021} proposed a method able to identify the specific frequency that characterizes a GAN engine through a deeper analysis of coefficients given from the Discrete Cosine Transform (DCT). These traces are characterized by both the network architecture (number and type of layers) and its specific parameters \cite{Yu2019AttributingFI}. 
Based on this principle, the synthetic images created by various GAN engines are also characterized by different statistics in terms of correlations between pixels. To capture this trace left by the convolutional layers, Guarnera et al.~\cite{guarnera2020Deepfake,guarnera2020fighting} proposed a method based on the Expectation-Maximization~\cite{moon1996expectation} algorithm, obtaining excellent classification results in distinguishing pristine data from deepfakes. 
Wang et al.~\cite{wang2020cnn} proposed a method to discriminate real images from those generated by ProGAN~\cite{karras2018progressive}. The method turns out to be able to generalize with synthetic data created by different GAN architectures. 

Recent solutions use Vision Transformer to detect deepfakes~\cite{coccomini2022combining,heo2023deepfake}. For example, \cite{wodajo2021Deepfake} combined vision transformers with a convolutional network, achieving excellent results in solving the proposed task.

Researchers are also actively engaged in developing advanced techniques to identify synthetic images generated by Diffusion Models~\cite{sohl2015deep}. Corvi et al.~\cite{corvi2023detection} investigated the challenges associated with distinguishing synthetic images produced by diffusion models from authentic ones. They assess the suitability of current state-of-the-art detectors for this specific task.
Sha et al.~\cite{sha2022fake} proposed DE-FAKE, a machine-learning classifier-based method designed for the detection of diffusion models on four prevalent text-image architectures. Meanwhile, Guarnera et al.~\cite{guarnera2024mastering} introduced a hierarchical approach based on recent architectures. This approach involves three levels of analysis: determines whether the image is real or manipulated by any generative architecture (AI-generated); identifies the specific framework, such as GAN or DM; defines the specific generative architecture among a predefined set.

Experimental results of all these methods have demonstrated that generative models leave unique traces that can be detected to distinguish deepfakes well from real multimedia content.

\subsection{Adversarial Attacks}
Adversarial attack methods are designed to introduce imperceptible changes to images with the aim of disrupting the feature extraction process performed by neural networks. Initially applied in classification tasks~\cite{goodfellow2014explaining}, ~\cite{moosavi2016deepfool}, \cite{dong2018boosting}, where their goal was to induce misclassification errors, these methods have been extended to segmentation~\cite{fischer2017adversarial} and detection tasks~\cite{xie2017adversarial}. However, the optimization process of unique pattern for each individual image can be highly time-consuming. To address this challenge, researchers introduced the concept of generic universal image-agnostic noise patterns~\cite{hendrik2017universal},~\cite{moosavi2017universal}. Such noise patterns are designed to be versatile and applicable across a wide range of images, eliminating the need for time-consuming, image-specific pattern optimization. While this approach has proven effective in the context of tasks involving misclassification, it has demonstrated limitations when applied to generative models.

\subsection{Image Manipulation Prevention}
Prevent image manipulations exploiting adversarial attack techniques has been recently studied as an alternative way to the classification and detection of manipulated images. The authors of~\cite{ruiz2020disrupting} propose a baseline methods for disrupting deepfakes by adapting adversarial attack methods to image translation networks. In~\cite{yeh2020disrupting},~\cite{yeh2021attack} the authors presented an approach to nullify the effect of image-to-image translation models. In~\cite{huang2021initiative} authors proposed a novel neural network based approach to generate image-specific patterns for low-resolution images which differs from the previous methods because does not require optimization of a specific pattern for each image separately which is computationally expensive. In~\cite{aneja2022tafim} an innovative framework called Targeted Adversarial Attacks for Facial Forgery Detection (TAFIM), a innovative framework that accepts a real image $X_i$ and a global perturbation $\delta_G$ as inputs to the model. This process generates an image-specific perturbation $\delta_i$. The resulting perturbation is then added to the original image, producing the protected image $X^p_i$, which is subsequently processed through the manipulation model $f_\phi$. The outcome is the manipulated output $Y^p_i$, utilized for driving the optimization process.

\section{Proposed Method}
\label{sec:proposed}
Our goal is to prevent image manipulation, specifically the addition or removal of tumors in CT scans, by disrupting the CT-GAN~\cite{mirsky2019ct} architecture. We designed a proper way to introduce an imperceptible perturbation that disrupts the CT-GAN's output in case of malicious manipulation, making it easier for a human to identify tampered scans, and hence ensuring the integrity of medical imaging process. 
MITS-GAN operates by applying protection at a slice-by-slice level for 3D CT scans. Rather than implementing a global protection mechanism across the entire 3D volume, our approach applies 2D convolutions to each slice independently. This localized protection ensures that even if only a subset of slices is manipulated, the algorithm remains robust, as each slice is protected individually. This slice-wise protection is particularly advantageous in scenarios where tampering occurs in specific areas of the scan, as it allows the detection of subtle and localized changes. In contrast to recent methods that use 3D-based GANs \cite{mirsky2019ct,shi20233dgaunet} to solve tasks such as creating new datasets or performing attacks on medical images, a slice-wise approach such as the one we propose, can offer greater advantages in terms of computational efficiency and flexibility in both the creation of new synthetic data and the handling of partial manipulations.

The chosen architecture leverages Generative Adversarial Networks (GANs) to generate protected images using a Gaussian perturbation (noise). The primary idea is to ensure that these protected images are indistinguishable from the original ones.
By concatenating the noise as an additional channel rather than directly adding it to the CT scan images ($x$), potential image artifacts are avoided. This approach helps the network treat the perturbation as extra information rather than as part of the image data itself, which could otherwise lead to unwanted artifacts that might be discarded during training.
The inclusion of Mean Squared Error (MSE) loss, which is maximized during training, plays a crucial role. This loss function compels the network to generate robust images that resist manipulations from the CT-GAN model, thereby preserving fidelity to the original images.

\subsection{Overview}

\begin{figure}[t!]
\centering
\includegraphics[width=1\textwidth]{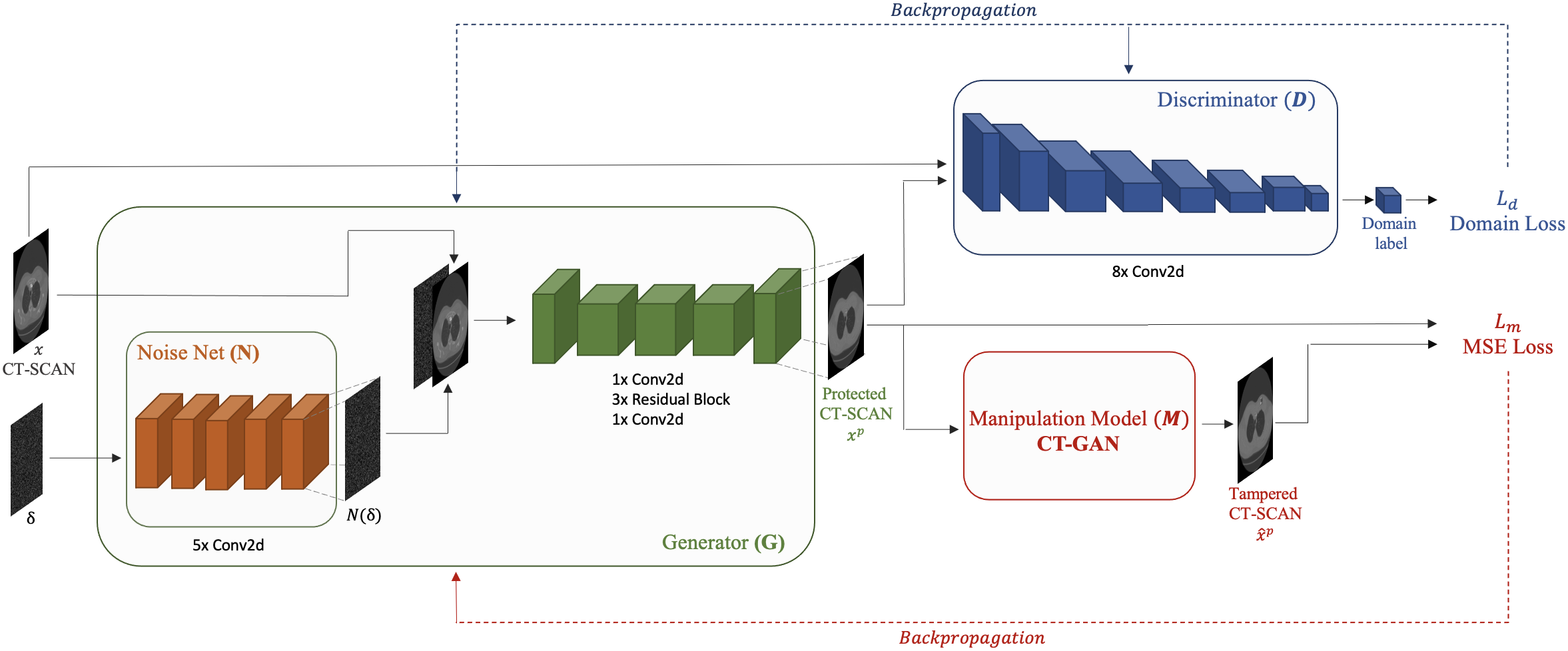}
\caption{Model Architecture Overview: The generator receives the input image $x$ and perturbation noise $\delta$ to produce the protected image $x^p$. Subsequently, $x^p$ is forwarded to the manipulation model and discriminator.}
\label{fig:protection_model}
\end{figure}

The proposed architecture is illustrated in Figure~\ref{fig:protection_model}. Let $\delta$ be an image-agnostic perturbation, distributed according to a Gaussian distribution, and $X = \{x_i\}^{N}_{i=0}$ be the  of $N$ CT scans. $\delta$ and $x$ are fed into the Generator $G(x,\delta;\theta_G)$ of parameters $\theta_G$, which includes a Noise Net N that, for a given input $\delta$, outputs $N(\delta)$. This module contains five 2D convolutional layers, each followed by batch normalization and the ReLU activation function. Subsequently, the CT scan $x$ and $N(\delta)$ are concatenated channel-wise and passed through a sequence of convolutional layers (one 2D convolution, three residual blocks, one 2D convolution). Each layer is followed by batch normalization and the ReLU activation function, except for the last one, which applies the Tanh activation function. Concatenating the noise as a new channel allows the network to consider the perturbation as extra information instead of adding it to $x$, which could lead to it being considered as image artifacts and therefore discarding them in the training phase to make the generator's output similar to $x$. The resulting output of $G$, denoted as $x^p$, represents the protected scan and is forwarded to the CT-GAN manipulation model $M$ which tamper the $x^p$ producing $\hat{x}^p$ and whose parameters are frozen. Additionally, $x$ and $x^p$ are provided to the discriminator $D(x;\theta_D)$ that outputs the likelihood $d$ that a given image $x$ belong to the real images with the aim to distinguish between a protected image produced by the generator and the original unprotected one. The Discriminator $D$ consists of eight 2D convolutional layers, each followed by batch normalization and the LeakyReLU activation function. The model is trained using a generative adversarial objective, encouraging the generator to produce protected images similar to the original (unprotected) ones.

The goal is to optimize the following min-max objective:
\begin{equation}
\min_{G}\max_{D,M} L_d(D,G) + \alpha L_m(G,M)
\end{equation}
where $L_d$ represents the domain loss:
\begin{equation}
    L_d(D,G) = \E_{x^p}[log D(x^p;\theta_D)] + \E_{x, \delta}[log(1 - D(G(x,\delta;\theta_G);\theta_D))]
\end{equation}
where $\E$ 
denotes the average value of the enclosed expression over the specified distribution. In detail, $\E_{x^p}[log D(x^p;\theta_D)]$ represents the expected log-probability that the discriminator assigns to real data samples. The discriminator aims to maximize this term, meaning it tries to correctly identify real data as real. $\E_{x, \delta}[log(1 - D(G(x,\delta;\theta_G);\theta_D))]$ represents the expected log-probability that the discriminator assigns to fake data samples created by the generator. The discriminator aims to maximize this term by correctly identifying fake data as fake (i.e., assigning a low probability to fake data being real).

$L_m$ is the Mean Squared Error (MSE) loss computed between the output of the model $M$ and the generator $G$:
\begin{equation}
    L_m(G,M) = \E_{x,\delta}[(M(G(x,\delta;\theta_G)) - G(x,\delta;\theta_G))^2]
\end{equation}
where $\E_{x,\delta}[(M(G(x,\delta;\theta_G)) - G(x,\delta;\theta_G))^2]$ denotes that the expectation is taken over the distributions of $x$ and $\delta$, indicating that we are considering the average squared error across all possible input and noise pairs.

$\alpha$ is the weight that controls the interaction of these losses. 
In particular, the optimization of the loss function $L_d$ concerning both the discriminator $D$ and the generator $G$ constitutes the standard generative adversarial objective. This objective concurrently refines both the generator and the discriminator. Subsequently, the term $L_m$ is introduced to augment the visual dissimilarity between the generated output $x^p$ and its corresponding tampered image $\hat{x}^p$. The inclusion of $L_m$ serves the purpose of increasing noticeable artifacts in the manipulated content $\hat{x}^p$ when attempting to tamper with $x^p$. Algorithm~1 reports the complete forward procedure of the proposed method. 
\section{Dataset, Metrics and Experimental Results}
\label{sec:dataset_metrics}

\begin{algorithm}[t!]
  \caption{Forward pass description of the proposed framework}
  \label{alg:forward_pass}
  \SetAlgoNlRelativeSize{0}
  
  \textbf{Input:} $CT \, \text{scan} = x$, $\delta = \text{perturbation}$\;
  
  \textbf{Step 1:} Forward $\delta$ and $x$ through the Generator $G$\;
  
  \textbf{Step 2:} Feed $\delta$ into the Noise Net $N$ within $G$, obtaining $N(\delta)$, and concatenate it with $x$\;
  
  \textbf{Step 3:} Apply five convolutions in $G$ to generate the protected image $x^p$\;
  
  \textbf{Step 4:} Pass $x^p$ to the Discriminator $D$ and the Manipulation model $M$ (CT-GAN)\;
  
  \textbf{Step 5:} Compute domain loss for $x$ and $x^p$ through $D$\;
  
  \textbf{Step 6:} Utilize $M$ to extract and manipulate a $32\times 32$ pixel square $q$. Generate a tampered image $\hat{x}^p$ by pasting $q$ onto $x^p$\;
  
  \textbf{Step 7:} Compute MSE loss between $x^p$ and $\hat{x}^p$\;
  
  \label{algorithm_label}
\end{algorithm}

In this section, we expound upon the dataset, outline the metrics under consideration, and scrutinize the outcomes derived from the introduced methodology. The manipulation model, denoted as CT-GAN, operates by taking a CT scan as input, identifying a designated square for manipulation, and subsequently producing the manipulated square. This process involves either removing or adding a tumor, resulting in the tampered square, which is then seamlessly integrated into the original scan. It is noteworthy that the tampered scan closely resembles the original, with the sole exception being the generated square. Our model ensures the comprehensive protection of the entire scan, as manipulations can be applied to any part of the scan, necessitating robustness across the entire image.
\subsection{Dataset}
\begin{figure}[t!]
\centering
\includegraphics[width=0.7\textwidth]{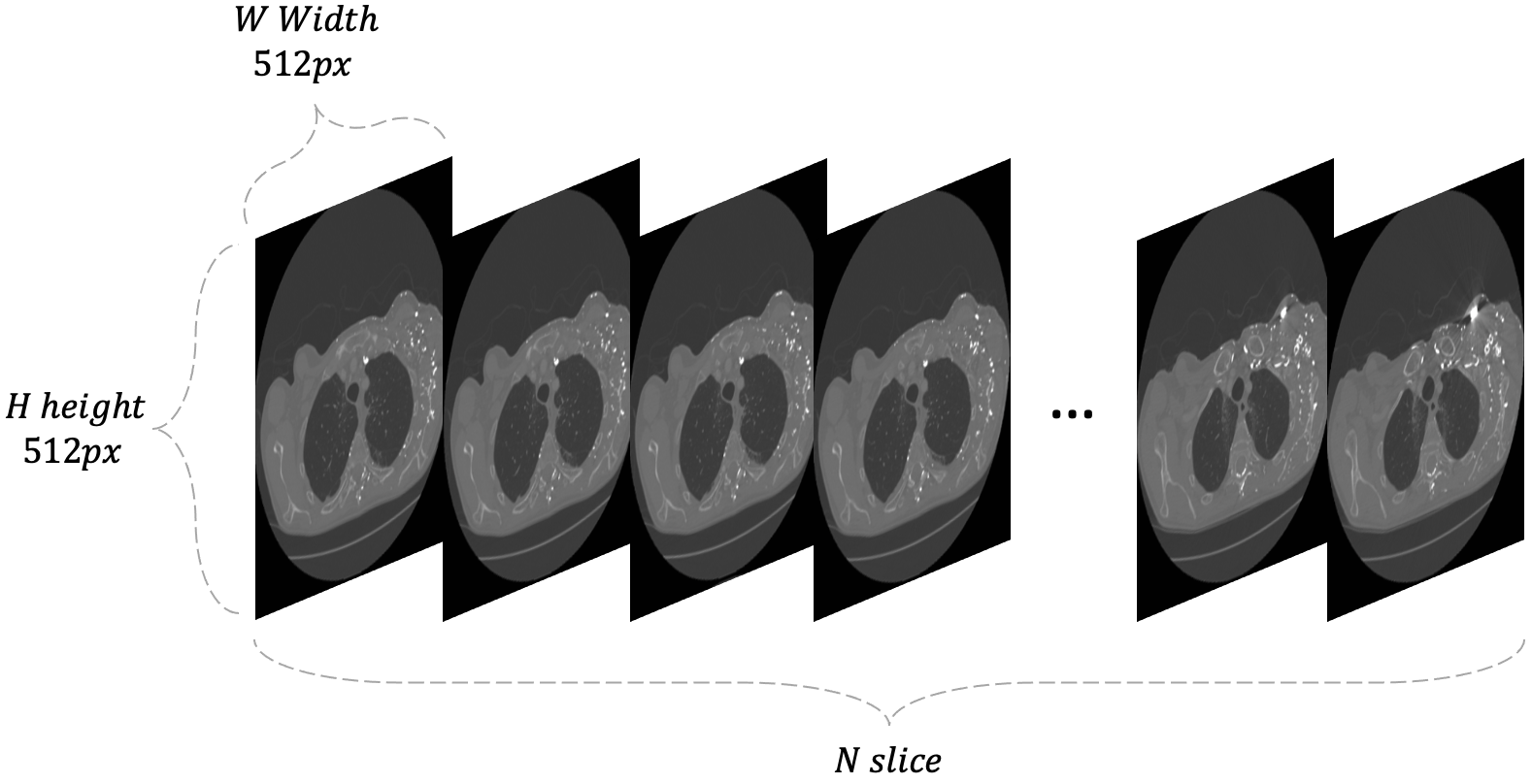}
\caption{Example of a CT scan.}
\label{fig:dataset}
\end{figure}
Our approach is evaluated using the dataset outlined in~\cite{armato2011lung}, following the training editing procedure specified in~\cite{mirsky2019ct}. In this procedure, the authors' injector model is trained on cancer samples with a minimum diameter of 10 mm, while the remover model is trained on benign lung nodules with a diameter less than 3 mm. The dataset comprises 888 CT scans, and we adhered to the standard split procedure, allocating 80\% as the training set and the remaining 20\% as the test set. Each CT scan is stored as a DICOM or Raw file, and its dimensions are represented as $N\times H\times W$, where $N$ identify the number of ``slices" or thin sections through which the scan was performed, $H$ represents the height, and $W$ represents the width of the scan (see Figure~\ref{fig:dataset}). The considered CT scans have a fixed resolution of $512\times512$ and a variable number of slices within the range $N \in [95, 764]$.

\subsection{Metrics}
To evaluate the output quality in a quantitative way, we compute the RMSE, PSNR, LPIPS~\cite{zhang2018unreasonable} and SSIM metrics as detailed below:
\begin{itemize}
    \item RMSE (Root Mean Square Error) measure the deviation between predicted values from a model and the actual observed values. Lower values are better, 0  zero indicates that the predicted values are equals to the observed values.
    \begin{equation}
        RMSE = \sqrt{\frac{1}{N}\sum^N_{i=1} || y_i - x_i||^2}
    \end{equation}
    \item PSNR (Peak signal-to-noise ratio) is a metric used to quantify the quality of an image or video by measuring the ratio of the maximum possible signal strength to the noise introduced during compression or transmission. Higher PSNR values generally indicate better image quality. 
    \begin{equation}
        PSNR(I, J) = 10 \cdot \log_{10}\left(\frac{MAX_I^2}{MSE}\right) \quad MSE = \frac{1}{N}\sum^N_{i=1} (y_i - x_i)^2
    \end{equation}

    \item LPIPS computes the similarity between the feature representations of two image patches extracted by a pre-trained neural network. This metric has demonstrated a strong alignment with human perception. The lower the LPIPS score, the more perceptually similar the image patches are considered to be. For the experiment we used as pretrained network SqueezeNet~\cite{iandola2016squeezenet}.

    \item SSIM computes the similarity between two images based on their structural similarity, taking into account factors such as luminance, contrast, and structural patterns. Higher SSIM values indicate greater similarity between the two images according to human visual perception.
    \begin{equation}
        \text{SSIM}(x, y) = \frac{(2\mu_x\mu_y + C_1)(2\sigma_{xy} + C_2)}{(\mu_x^2 + \mu_y^2 + C_1)(\sigma_x^2 + \sigma_y^2 + C_2)}
    \end{equation}
    $\mu_x$ and $\mu_y$ are the mean intensities of images $x$ and $y$.
    $\sigma_x$ and $\sigma_y$ are the standard deviations of images $x$ and $y$.
    $\sigma_{xy}$ is the covariance between $x$ and $y$.
    $C_1$ and $C_2$ are small constants to stabilize the division with weak denominator.

\end{itemize}

\subsection{Experimental Setup}
All models were trained for 20 epochs using a NVIDIA V100. The MITS-GAN~\footnote{https://github.com/GiovanniPasq/MITS-GAN} architecture, implemented using PyTorch~\footnote{https://pytorch.org/}, was trained with a batch size of 16, a learning rate set at 0.0002, betas of $[0.5,0.999]$, and utilizing Adam as the optimizer. For TAFIM, we adopted the configurations suggested by the authors in~\cite{aneja2022tafim}.

\subsection{Results}
\begin{figure}[t!]
\centering
\includegraphics[width=1\textwidth]{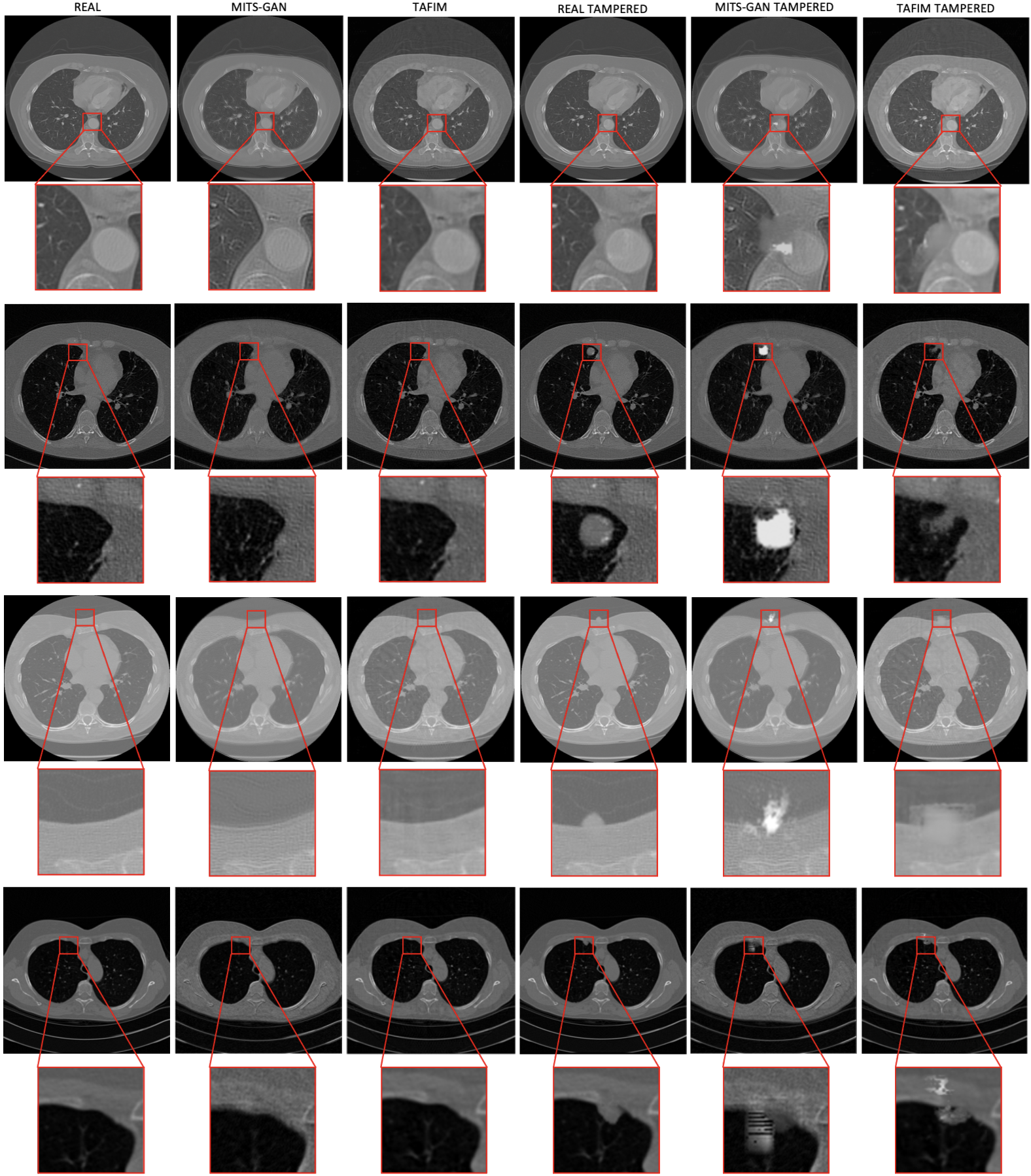}
\caption{Qualitative results on the reconstruction task compared with images as manipulation targets.}
\label{fig:qualitative_results_medical}
\end{figure}

\begin{figure}[t!]
\centering
\includegraphics[width=0.9\textwidth]{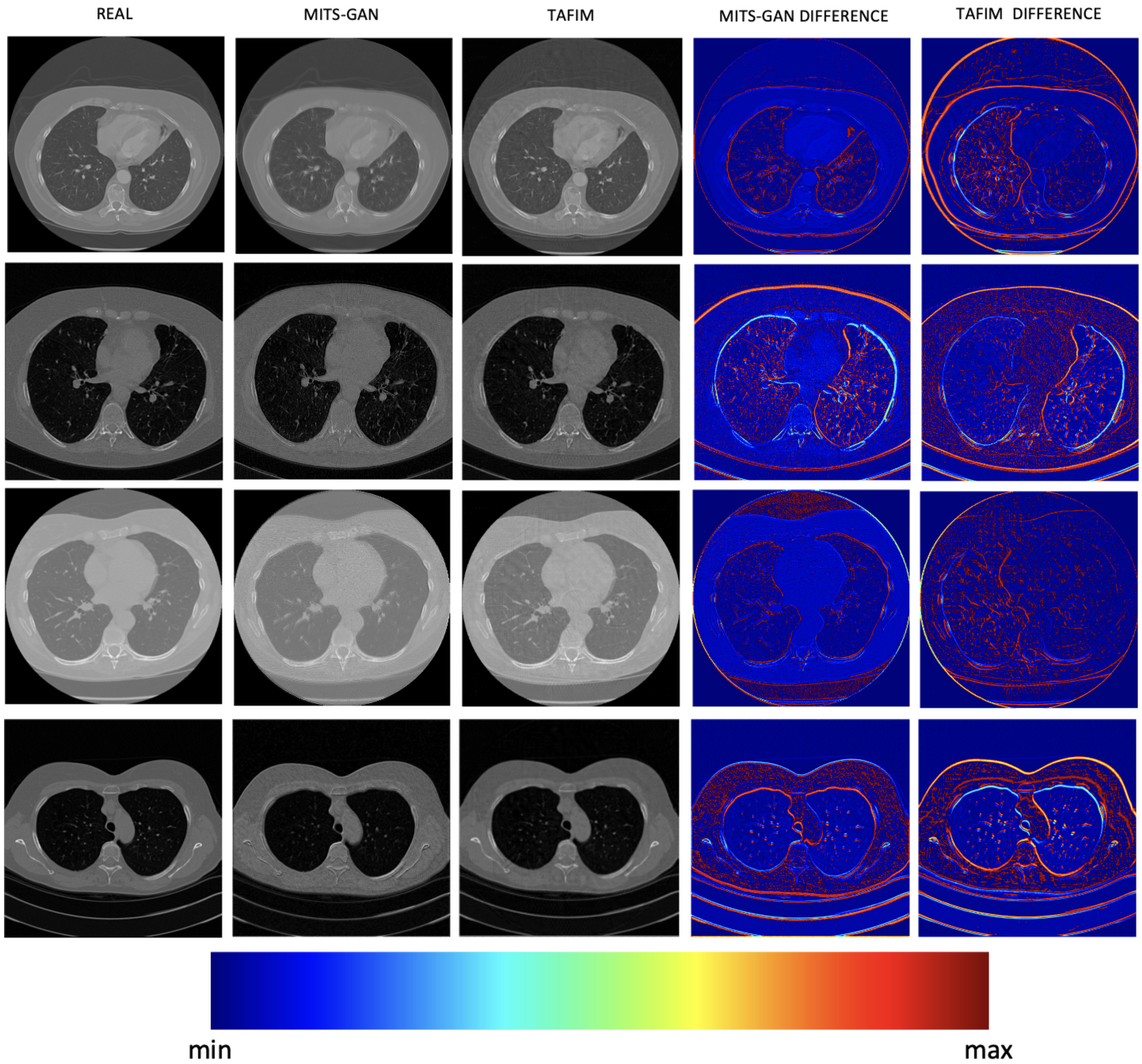}
\caption{Heatmap computed between the pairs real-MITS-GAN and real-TAFIM.}
\label{fig:qualitative_results_medical2}
\end{figure}

Figure~\ref{fig:qualitative_results_medical} shows the qualitative results of the proposed MITS-GAN method compared with TAFIM~\cite{aneja2022tafim}. MITS-GAN exhibits fewer visible artifacts on the reconstructed images and demonstrates a more robust ability to resist manipulation, accentuating the artifacts introduced when the model attempts to manipulate the selected square. Figure~\ref{fig:qualitative_results_medical2} shows the heatmap obtained by performing a pixel-to-pixel difference between the real image and the protected one. In this case, the proposed method generates protected images that are more faithful to the originals than the compared method.

\begin{table}[t!]
\caption{Metric results evaluated between the following pairs on the: real-MITS-GAN, real-TAFIM, real-MITS-GAN tampered and real-TAFIM tampered. Lower values are better for RMSE and LPIPS, higher for PSNR and SSIM.}
\label{table1_medical}
\centering
\begin{tabular}{ccc||cc}
\cline{2-5}
                & \multicolumn{2}{c}{\textbf{Real}}  & \multicolumn{2}{c}{\textbf{Tampered}}    \\ \hline
\textbf{Metric} & \textbf{MITS-GAN} & \textbf{TAFIM} & \textbf{MITS-GAN T.} & \textbf{TAFIM T.} \\ \hline
\textbf{RMSE}           & 169.481           & 194.943        & 198.253              & 233.780           \\ \hline
\textbf{PSNR}          & 27.949            & 21.702         & 21.237               & 21.469            \\ \hline
\textbf{LPIPS}           & 0.170             & 0.383          & 0.226                & 0.391             \\ \hline
\textbf{SSIM}          & 0.983             & 0.945          & 0.970                & 0.981             \\ \hline
\end{tabular}
\end{table}

\begin{table}[t!]
\caption{Metric results evaluated between the following pairs on the tampered square part of the images: real-MITS-GAN, real-TAFIM, real-MITS-GAN tampered and real-TAFIM tampered. Lower values are better for RMSE and LPIPS, higher for PSNR and SSIM.}
\label{table2_medical}
\centering
\begin{tabular}{ccc||cc}
\cline{2-5}
                & \multicolumn{2}{c}{\textbf{Real}}  & \multicolumn{2}{c}{\textbf{Tampered}}    \\ \hline
\textbf{Metric} & \textbf{MITS-GAN} & \textbf{TAFIM} & \textbf{MITS-GAN T.} & \textbf{TAFIM T.} \\ \hline
\textbf{RMSE}   & 50.565            & 66.061         & 84.349               & 79.451            \\ \hline
\textbf{PSNR}   & 26.682            & 18.854         & 11.289               & 18.511            \\ \hline
\textbf{LPIPS}  & 0.372             & 0.3417         & 0.591                & 0.346             \\ \hline
\textbf{SSIM}   & 0.992             & 0.972       & 0.740          & 0.866             \\ 
\hline
\end{tabular}
\end{table}
Table~\ref{table1_medical} reports the results of the considered metrics evaluated between each pair of real-protected and real-protected/tampered on the entire images. MITS-GAN has lower RMSE, LPIPS, and higher PSNR and SSIM values compared to TAFIM, suggesting better reconstruction quality of the images. This advantage is maintained even when considering the images after manipulation. Table~\ref{table2_medical} shows the results evaluated on the square part subjected to manipulation. In this case, the metrics favor the proposed method. After manipulation, the output produced by the manipulator model appears to be more damaged than the compared method. This suggests that MITS-GAN produces images with less noise but is more robust to manipulation, generating more visible artifacts when attempting to tamper with an image.

\subsection{Ablation Study}
Table~\ref{table_ablation} presents the results of MITS-GAN varying the hyperparameter $\alpha$, which regulates the standard GAN losses and the MSE loss used to generate robust images against manipulation by CT-GAN. Since CT-GAN performs manipulation on a square of size $32\times32$ pixels, the evaluation considers which $\alpha$ value provides the best protection. This assessment focuses on maximizing RMSE and LPIPS while minimizing PSNR and SSIM. The goal is to ensure that the output generated by CT-GAN after manipulation is significantly different from the original, introducing artifacts that are clearly visible to the human eye. As shown in the table, the best performance is achieved when $\alpha = 1$.

\begin{table}[t!]
\caption{Ablation study about the impact of the MSE loss.}
\label{table_ablation}
\centering
\begin{tabular}{cccccc}
\hline
\multirow{2}{*}{\textbf{Metric}} & \multicolumn{5}{c}{\textbf{$\alpha$}}                                      \\
                                 & \textbf{0.2} & \textbf{0.4} & \textbf{0.6} & \textbf{0.8} & \textbf{1} \\ \hline
\textbf{RMSE}                    & 79.026       & 80.472       & 81.920       & 82.517       & 84.349     \\ \hline
\textbf{PSNR}                    & 18.766       & 17.145       & 15.803       & 13.562       & 11.289     \\ \hline
\textbf{LPIPS}                   & 0.338        & 0.377        & 0.425        & 0.510        & 0.591      \\ \hline
\textbf{SSIM}                    & 0.881        & 0.854        & 0.810        & 0.775        & 0.740      \\ \hline
\end{tabular}
\end{table} 

\section{Discussion}
\label{sec:discussion}

The MITS-GAN approach has shown considerable promise in safeguarding medical imaging from tampering, particularly when compared to existing methods such as TAFIM. Experimental results show that MITS-GAN achieves lower RMSE and LPIPS values and higher PSNR and SSIM values, indicating superior image reconstruction quality and robustness against manipulation. 
MITS-GAN succeeds in creating high-quality images that are almost completely identical to the originals and with almost no artifacts. This robustness is crucial in medical imaging where clarity and accuracy are fundamental. In addition, the method generates tamper-resistant images, showing more visible artifacts when data protected by MITS-GAN is tampered with by other architectures, making it easier to detect non-authorized alterations.
Despite these strengths, some limitations and potential areas for improvement can be identified:

\begin{itemize}
    \item \textit{Sensitivity of hyper-parameters}: MITS-GAN's performance largely depends on the careful tuning of hyper-parameters, such as the $\alpha$-value that balances GAN loss and MSE loss. Incorrect tuning can have a significant impact on the effectiveness of the model.
    \item \textit{Computational complexity}: MITS-GAN training requires some computational resources, including high-performance GPUs and extended training times, which may limit its accessibility and implementation in resource-limited environments.
\end{itemize}
It is important to note that the computational problems (in terms of time) are mainly related to the model training procedure.
The protection of CT scans using the MITS-GAN method does not require heavy computational resources because real-time protection during acquisition is unnecessary. This approach allows for protection to be performed later, in the background, without impacting the primary image acquisition process. Therefore, from the standpoint of scalability and computational efficiency, MITS-GAN proves suitable for practical applications in the medical domain, enabling efficient resource management without compromising service quality.

Future works will focus on improving the MITS-GAN architecture considering:

\begin{itemize}
    \item \textit{Integration of Diffusion Models}
            One promising direction involves integrating diffusion models into the MITS-GAN framework. Diffusion models, known for iteratively adding noise to images, could contribute to improving the quality and authenticity of safeguarded medical imagery generated by MITS-GAN.
            
    \item \textit{Attention Mechanisms for Robustness}
            To fortify MITS-GAN against malicious tampering, future work could incorporate attention mechanisms. Attention mechanisms enable the model to focus on relevant regions of the input, potentially making it more resilient to adversarial attacks and ensuring critical details in medical scans are preserved.
            
    \item \textit{Exploring Diverse Architectures}
            The success of MITS-GAN opens the door to exploring diverse generative model architectures. Investigating different GAN variants or hybrid architectures could provide valuable insights into optimizing the trade-off between image quality, computational efficiency, and security.

    \item \textit{Real-world Deployment and Validation}
            A crucial step toward practical application involves focusing on real-world deployment and validation of MITS-GAN. Collaborations with healthcare institutions and professionals can provide valuable feedback, ensuring that the proposed method aligns with the practical requirements and standards of the medical imaging community.
\end{itemize}

\section{Conclusion}
\label{sec:conclusion}
In this work, we introduced MITS-GAN, an innovative approach to safeguard medical imagery against malicious tampering. The method demonstrated superior performance in disrupting manipulations at the source, resulting in the generation of tamper-resistant images with fewer artifacts when compared to existing techniques. The proactive measures outlined in this study hold significant importance in guaranteeing the responsible and ethical use of generative models, particularly in critical applications such as healthcare.
By addressing the vulnerabilities in medical imaging systems, MITS-GAN contributes to the overall resilience of these systems against potential threats.
Looking ahead, future works and potential extensions aim to further refine and enhance the capabilities of MITS-GAN. This ongoing research aligns with our commitment to staying at the forefront of advancements in securing medical imaging technology. By continually pushing the boundaries of innovation, we aim to make meaningful contributions that strengthen the integrity and reliability of healthcare systems, and ensuring the trustworthiness of medical diagnostic tools.

\section*{Acknowledgments}
This research is supported by research Program PIAno di inCEntivi per la Ricerca di Ateneo 2020/2022 — Linea di Intervento 3 "Starting Grant" - University of Catania, Italy.
This research is supported by Azione IV.4 - ``Dottorati e contratti di ricerca su  tematiche dell’innovazione" del nuovo Asse IV del PON Ricerca e Innovazione 2014-2020 “Istruzione e ricerca  per il recupero - REACT-EU”- CUP: E65F21002580005.

 \bibliography{main}

\end{document}